\index{\begin{small}\begin{normalsize}\label{\index{\label{\index{\footnote{\index{\footnote{\index{\index{\index{\footnote{\index{\footnote{\footnote{}}}}}}}}}}}} }}                     \end{normalsize}       \end{small}}
\documentclass[aps,prl,twocolumn,showpacs,groupedaddress]{revtex4}
\usepackage{graphicx}  
\usepackage{dcolumn}   
\usepackage{bm}        
\usepackage{amssymb}   
\begin{document}


\def\ra{{\rightarrow}}
\def\a{{\alpha}}
\def\b{{\beta}}
\def\l{{\lambda}}
\def\eps{{\epsilon}}
\def\T{{\Theta}}
\def\t{{\theta}}
\def\co{{\cal O}}
\def\car{{\cal R}}
\def\caf{{\cal F}}
\def\cs{{\Theta_S}}
\def\pr{{\partial}}
\def\tri{{\triangle}}
\def\na{{\nabla }}
\def\S{{\Sigma}}
\def\s{{\sigma}}
\def\sp{\vspace{.15in}}
\def\hs{\hspace{.25in}}

\newcommand{\be}{\begin{equation}} \newcommand{\ee}{\end{equation}}
\newcommand{\bea}{\begin{eqnarray}}\newcommand{\eea}
{\end{eqnarray}}

\title{\Large\bf Emergent Schwarzschild and Reissner-Nordstrom Black Holes in 4D:\\ An effective curvature sourced by a $B_2$-field on a $D_4$-brane}
\author{Abhishek K. Singh, K. Priyabrat Pandey, Sunita Singh and Supriya Kar}
\affiliation{Department of Physics \& Astrophysics, University of Delhi, New Delhi 110 007, India}
\begin{abstract}
\noindent
We obtain a Schwarzschild and a Reissner-Nordstrom emergent black holes, by exploring the torsion dynamics in a generalized curvature formulation, underlying an effective $D_4$-brane on $S^1$. It is shown that a constant effective metric, sourced by a background fluctuation in $B_2$-potential, 
on a $D_3$-brane receives a dynamical quantum correction in presence of an electric charge.
\end{abstract}
\maketitle
\section{Introduction}
\noindent 
The holographic idea underlying AdS/CFT duality \cite{maldacena,witten} is believed to be a powerful tool to revisit the gauge theories on a $D$-brane and its near horizon black hole geometries. In the past, there have been attempts to construct various near horizon $D$-brane effective geometries \cite{gibbons}-\cite{zhang3} by exploiting a non-linear electromagnetic field \cite{seiberg-witten}. For a recent review on brane-world gravity, see ref.\cite{maartens}.

\sp
\noindent
In the context, it may be thought provoking to explore the idea of gravity as an ``emergent'' phenomenon \cite{carlip} especially 
underlying a brane-world. Interestingly, a $U(1)$ gauge theory underlying a two form dynamics on a $D_4$-brane leading to a generalized curvature in a second order formalism has been addressed by the authors \cite{kpss1}-\cite{D4-torsion}. The generalized curvature has been shown to describe a propagating geometric torsion and various emerging brane geometries underlying de Sitter and AdS vacua have been obtained. Very recently, an exact Reissner-Nordstrom black hole on a brane in a different context has been anlyzed \cite{kiley}.

\sp
\noindent
In the article, we explore the generalized curvature in an effective $D_4$-brane on $S^1$ to obtain a Schwarzschild and a Reissner-Nordstrom emergent black holes in a $U(1)$ gauge theory. It is shown that the emergent geometries on an effective $D_3$-brane in certain windows identify themselves with the black hole geometries otherwise obtained in Einstein's gravity. The primary motivation behind the work lies in the emergent notion of gravity on an effective $D_3$-brane, essentially, sourced by a two form in a $U(1)$ gauge theory defined on a $D_4$-brane. In particular, it is argued that a generic torsion, intrinsic to the frame-work on an effective $D_4$-brane, vanishes in a gauge choice to yield some of the established black hole geometries in $4D$ underlying Einstein's gravity. Interestingly, the emergent brane geometries obtained in the article have been shown to incorporate quantum corrections, due to the $U(1)$ charges, into the background (metric) fluctuations in the formalism. It is shown that the quantum corrections may seen to be associated with a flat space-time metric. Our analysis attempts to connect two a priori different black hole geometries underlying two different formulations $i.e.$ gauge theoretic and Einstein's gravity, respectively, underlying a two form and a metric tensor. It may provoke thought to believe gravity as an ``emergent phenomenon''.

\section{Curvature on an effective ${\mathbf{D_4}}$-brane}
We begin with a $U(1)$ gauge theory on a $D_4$-brane. In presence of a constant background metric $g_{\mu\nu}$, the $A_{\mu}$-field dynamics is given by
\be
S= -{1\over{4C_1^2}}\int d^5x\ {\sqrt{-g}}\ F_{\alpha\beta}F^{\alpha\beta}\ ,\label{gauge-1}
\ee 
where $C_1^2=(4\pi^2g_s){\alpha'}^{1/2}$. Alternately, the Poincare dual to the electromagnetic field is described by a $B_2$-field which in turn describes a torsion. For instance, see ref.\cite{kubiznak}. The two form gauge dynamics is given by
\be
S=- {1\over{12C_2^2}}\int d^5x\ {\sqrt{-g}}\ H_{\mu\nu\lambda}H^{\mu\nu\lambda}\ ,\label{gauge-2}
\ee
where $C_2^2=(8\pi^3g_s){\alpha'}^{3/2}$ and $H_{\mu\nu\lambda}=3\nabla_{[\mu}B_{\nu\lambda ]}$. In the context, an appropriate covariant derivative ${\cal D}_{\mu}$ on a brane may be constructed using the two form connections. It modifies $H_3$ to ${\cal H}_3$ and incorporates an effective curvature into the brane gauge dynamics \cite{D4-torsion}. The covariant derivative in a gauge theory becomes
\be
D_{\lambda}B_{\mu\nu}=\nabla_{\lambda}B_{\mu\nu} - {{\mathbf\Gamma}_{\lambda\mu}}^{\rho}B_{\rho\nu} + {{\mathbf\Gamma}_{\lambda\nu}}^{\rho}B_{\rho\mu}\ ,\label{gauge-3}
\ee
where $-2{{\mathbf\Gamma}_{\mu\nu}}^{\rho}={H_{\mu\nu}}^{\rho}$. An iterative incorporation of $B_2$-corrections, to all orders, in the covariant derivative leads to an exact derivative in a perturbative gauge theory. It may seen to define a non-perturbative covariant derivative in a second order formalism underlying a geometric realization. It is given by
\bea
{\cal D}_{\lambda}B_{\mu\nu}&=&\nabla_{\lambda}B_{\mu\nu} + {1\over2}{{{\cal H}}_{\lambda\mu}}^{\rho}B_{\rho\nu} - {1\over2}{{\cal H}_{\lambda\nu}}^{\rho}B_{\rho\mu}\ ,\nonumber\\
{\rm where}&&\;\; {\cal H}_{\mu\nu\lambda}=H_{\mu\nu\lambda} + 3{{\cal H}_{[\mu\nu}}^{\alpha}
{B^{\beta}}_{\lambda ]}\ g_{\alpha\beta}\ .\label{gauge-4}
\eea
Interestingly, a $B_2$-field dynamics in a first order formalism may be viewed as a torsion dynamics on an effective $D_4$-brane in a second order formalism \cite{D4-torsion} leading to a fourth order generalized tensor:
\be
4{{\cal K}_{\mu\nu\lambda}}^{\rho}= 2\partial_{\mu}{{\cal H}_{\nu\lambda}}^{\rho} -2\partial_{\nu} {{\cal H}_{\mu\lambda}}^{\rho} + {{\cal H}_{\mu\lambda}}^{\sigma}{{\cal H}_{\nu\sigma}}^{\rho}-{{\cal H}_{\nu\lambda}}^{\sigma}{{\cal H}_{\mu\sigma}}^{\rho}.\label{gauge-5}
\ee
The generalized tensor is antisymmetric under an exchange of indices within a pair and is not symmetric under an exchange of its first pair of indices with the second. Hence, it differs from the Riemannian tensor $R_{\mu\nu\lambda\rho}$. However for a non-propagating torsion, 
${\cal K}_{\mu\nu\lambda\rho}\rightarrow R_{\mu\nu\lambda\rho}$. In a second order formalism the ${\cal H}_3$ dynamics on an effective $D_4$-brane may be approximated by
\be
S_{\rm D_4}^{\rm eff}= {1\over{3C_2^2}}\int d^5x {\sqrt{-{\tilde G}}}\ {\cal K}^{(5)}\ ,\label{gauge-6}
\ee
where ${\tilde G}=\det {\tilde G}_{\mu\nu}$. Generically the emergent metric takes a form 
$${\tilde G}_{\mu\nu}=\left ( {\tilde g}_{\mu\nu} + C\ {\bar{\cal H}}_{\mu\lambda\rho}{{\cal H}^{\lambda\rho}}_{\nu}\right )\ ,$$
where ${\tilde g}_{\mu\nu}=\left ( g_{\mu\nu} - B_{\mu\lambda}{B^{\lambda}{}}_{\nu}\right )$ and $B_{\mu\nu}$ signify the background fluctuations. The cosmological constant ${\tilde\Lambda}$, in the geometric action, is sourced by a background $B_2$-fluctuations in the frame-work. It is important to note that the generalized action (\ref{gauge-6}) describes the propagation of a geometric torsion ${\cal H}_3$. In an emergent scenario, a generalized metric tensor is constructed from the gauge fields in the formalism. With $\kappa^2=(2\pi)^{5/2}g_s\alpha'$, a generalized curvature dynamics on $S^1$ reduces to yield
\be
S_{\rm D_3}^{\rm eff}= {1\over{3\kappa^2}}\int d^4x {\sqrt{-G}}\ \left ( {\cal K}^{(4)}\ -\ {3\over4} {\bar{\cal F}}_{\mu\nu}
{\cal F}^{\mu\nu} \right )\ .\label{gauge-7}
\ee
$${\rm where}\qquad {\bar{\cal F}_{\mu\nu}}=(2\pi\alpha')\left (F_{\mu\nu} + {\cal H}_{\mu\nu}^{\lambda}{\cal A}_{\lambda}\right )\ .$$
The curvature scalar ${\cal K}^{(4)}$ is essentially sourced by a dynamical two form potential in a first order formalism. Its field strength is appropriately modified $H_3\rightarrow {\cal H}_3$ to describe a propagating torsion in four dimensions underlying a second order formalism. The fact that a torsion is dual to an axion (scalar) on an effective $D_3$-brane, ensures one degree of freedom. In addition, ${\cal F}_{\mu\nu}$ describes a geometric one form field with two local degrees on an effective $D_3$-brane. A precise match among the (three) local degrees of torsion in ${\cal K}^{(5)}$ on $S^1$ with that in ${\cal K}^{(4)}$ and ${\cal F}_{\mu\nu}$ reassure the absence of a dynamical dilaton field in the frame-work. The result is consistent with the fact that a two form on $S^1$ does not generate a dilaton field. Alternately an emergent metric tensor ${\tilde G}_{\mu\nu}$, essentially sourced by a dynamical two form, does not formally evolve under a compactification of an underlying $U(1)$ gauge theory (\ref{gauge-6}) on $S^1$. In addition the background metric $g_{\mu\nu}$, being non-dynamical on a $D_4$-brane, can not generate a dynamical scalar field or dilaton when compactified on $S^1$ in the frame-work.

\sp
\noindent
On the other hand, the energy-momentum tensor $T_{\mu\nu}$ is computed in a gauge choice: $${3\over4}{\bar{\cal F}}_{\mu\nu}{\cal F}^{\mu\nu}= {3\over{\pi\alpha'}} + {\cal K}^{(4)}\ .$$ 
Explicitly
\be
(2\pi\alpha')T_{\mu\nu}=\left ( {\tilde g}_{\mu\nu} + {\tilde C}\ {\bar{\cal F}}_{\mu\lambda}
{{\bar{\cal F}}^{\lambda}{}}_{\nu}+ C\ {\bar{\cal H}}_{\mu\lambda\rho}{{\cal H}^{\lambda\rho}{}}_{\nu}\right )\ .\label{gauge-8}
\ee
The gauge choice ensures that the $T_{\mu\nu}$ sources a nontrivial emergent geometry underlying a non-linear $U(1)$ gauge theory. Interestingly, the $T_{\mu\nu}$ sources an effective metric $G_{\mu\nu}$ in the frame-work. 
$T_{\mu\nu}$, with ($C={3\over4}$, ${\tilde C}={3\over2}$) and ($C=-{5\over4}$, ${\tilde C}=-{1\over2}$), respectively correspond to a $(+)$ve sign and a $(-)$ve sign in $G_{\mu\nu}$. A priori, they describe two inequivalent black holes on an effective $D_3$-brane. Two solutions for an emergent metric tensor is a choice keeping the generality in mind. Primarily, they dictate the quantum geometric corrections to a background metric.
Generically, the effective metric on a $D_p$-brane for $p\neq 4$ may be given by
\be
G_{\mu\nu}=\left ( g_{\mu\nu} - B_{\mu\lambda}{B^{\lambda}{}}_{\nu} \pm {\bar{\cal F}}_{\mu\lambda}{{\bar{\cal F}}^{\lambda}{}}_{\nu} \pm 
{\bar{\cal H}}_{\mu\lambda\rho}{{\cal H}^{\lambda\rho}{}}_{\nu} \right )\ .\label{gauge-9}
\ee
In a gauge choice, the ansatz for the gauge fields  on a $D_3$-brane, may be expressed as:
\bea
B_{t\theta}= B_{r\theta}={b\over{\sqrt{2\pi\alpha'}}}\; ,\  B_{\theta\phi}={p\over{\sqrt{2\pi\alpha'}}}\ \sin^2\theta,&&\nonumber\\
{\cal A}_t\rightarrow A_t={{Q_e}\over{r}}\ ,\;  {\cal A}_{\phi}\rightarrow A_{\phi} = - {{Q_m}\over{\sqrt{2\pi\alpha'}}}\cos \theta,&&\label{gauge-8}
\eea 
where ($b,p$)$>$$0$ are constants. The gauge choice re-assures a vanishing torsion $H_3=0={\cal H}_3$, in presence of the $B_2$-fluctuations. Importantly, it leads to a four dimensional Einstein's gravity in the frame-work. It re-assures the generalized nature of an irreducible curvature scalar ${\cal K}^{(4)}$. In the gauge choice (\ref{gauge-8}), the emergent metric constructed on an effective $D_3$-brane reduces to yield
\be
G_{\mu\nu}\rightarrow\left ( g_{\mu\nu} - B_{\mu\lambda}{B^{\lambda}{}}_{\nu}\ \pm \ {\bar F}_{\mu\lambda}{{\bar F}^{\lambda}{}}_{\nu}\right )
\ .\label{gauge-10}
\ee
It is important to note that the local degrees in massless $A_{\mu}$ precisely source a dynamical $G_{\mu\nu}$ as required in four dimensional Einstein's gravity. In addition, the gauge field incorporates quantum corrections to the background metric ${\tilde{g}}_{\mu\nu}$. In the article, the effective dynamical $G_{\mu\nu}$ shall be shown to describe semi-classical geometries leading to a modified Schwarzschild and a Reissner-Nordstrom black hole.
\section{Quantum corrections to Schwarzschild and Asymptotic AdS}
In this section, we obtain the emergent brane geometries explicitly in term of their geometric quantum corrections. Interestingly, the corrections may seen to be associated with a flat metric, underlying a Minkowski space-time, in presence of the $U(1)$ charges. The line elements on an effective $D_3$-brane, in presence of a background $B_2$-fluctuations and an electromagnetic field, are worked out with a $S^2$-line element 
$d\Omega^2= \left ( d\theta^2 + \sin^2\theta\ d\phi^2 \right )$. It is is given  by
\bea
ds^2&=&-\left (1-{{b^2}\over{r^2}} \pm {{(2\pi\alpha'Q_e)^2}\over{r^4}}\right ) dt^2\nonumber\\
&+&\left ( 1 +{{b^2}\over{r^2}} \pm {{(2\pi\alpha'Q_e)^2}\over{r^4}}\right ) dr^2 +{{2b^2}\over{r^2}}\ dt dr\nonumber\\ 
&+&\left ( 1 + {{2\pi\alpha' p^2 \sin^2\theta}\over{r^4}}\mp {{(2\pi\alpha'Q_m)^2}\over{r^4}}\right )r^2 d\Omega^2\nonumber\\
&-&{{2(2\pi\alpha')^{1/2}bp \sin^2 \theta}\over{r^2}}(dt + dr)\ d\phi\ ,\label{RN-1}
\eea
In the limit $r$$>$$>$$b$, the emergent brane geometries may be simplified further. Redefining the parameters: $b=(2\pi\alpha')M\rightarrow M$, $(2\pi\alpha')Q_e\rightarrow Q_e$, $(2\pi\alpha')Q_m\rightarrow Q_m$ and $(2\pi\alpha')^{1/2}p\rightarrow p$, the brane geometries in the regime presumably identify themselves with some of the $4D$ black holes well established in Einstein's gravity. In addition, the emergent black hole may seen to receive quantum corrections due to a non-linear $U(1)$ charge in the frame-work. A non-zero $b$ is vital to the frame-work, which is essentially due to a $B_2$-field on a $D$-brane (\ref{gauge-8}). For a constant $B_2$, it may seen to describe a minimal length scale underlying a non-commutative parameter $\Theta$ on a $D$-brane \cite{seiberg-witten}. Thus the lower cut-off on $r$ is strictly defined on a brane, underlying a non-linear $U(1)$ gauge theory. It may not have any relevance to the vacuum solutions obtained otherwise in Einstein's gravity. The regime $r$$>$$>$$b$ on an emergent $D_3$-brane geometries may be approximated to yield
\bea
ds^2&=&-\left (1-{{M^2}\over{r^2}}\right ) dt^2+\left ( 1 -{{M^2}\over{r^2}}\right )^{-1} dr^2\nonumber\\
&+& {{2M^2}\over{r^2}}dt dr + r^2 d\Omega^2\ -\ {{2Mp \sin^2 \theta}\over{r^2}} (dt + dr)d\phi\nonumber\\
&\pm&{{Q_e^2}\over{r^4}}\left ( -dt^2 + dr^2 \pm {{p^2\sin^2\theta \mp Q_m^2}\over{Q_e^2}} r^2 d\Omega^2\right ).\label{RN-2}
\eea
A significant section, of the emergent geometries, is characterized by a single parameter $M$. Interestingly this section of emergent brane geometry 
may be identified with a classical Schwarzschild black hole presumably with a topological charge on an effective $D_3$-brane. The parameter $p$ in association with $M$, $i.e.\ (Mp)$, denotes an angular velocity intrinsic to a torsion in the formalism. The $B_2$-potential hints at an extra compact dimension which signifies a $D_4$-brane on $S^1$. In absence of the electric $Q_e$ and magnetic $Q_m$ charges, the emergent geometry on a brane is worked out for its invariant curvatures in Einstein's gravity. For functions $(f_1,f_2,f_3)$, the curvatures are given by

\vspace{-.15in}
$$R={{M^2}\over{r^2}}\ f_1(r,\theta,M,p)\ ,$$
\vspace{-.2in}
$$R_{\mu\nu}R^{\mu\nu}= {{M^2}\over{2r^4}}\ f_2(r,\theta,M,p)$$
\vspace{-.2in} 
$${\rm and}\quad R_{\mu\nu\lambda\rho}R^{\mu\nu\lambda\rho}= {{M^2}\over{2r^4}}\ f_3(r,\theta,M,p)\ .$$   

\noindent
They re-confirm a generic curvature singularity at $r\rightarrow 0$, which is identical to that in a typical Schwarzschild black hole in Einstein's gravity. Most importantly the curvature singularity at $r\rightarrow 0$ in Weyl-conformal tensor $C_{\mu\nu\lambda\rho}C^{\mu\nu\lambda\rho}$ in the emergent gravity scenario as well as in Einstein's gravity is remarkable. However the curvature singularity is not accessible to an observer on a brane-world due to the lower cut-off value $M$, on $r$, enforced by a $B_2$-field in the formalism. It implies that an emergent Schwarzschild black hole on a brane is always covered by a horizon at $r_S$$=$$M$. The $U(1)$ charges in eq(\ref{RN-2}) incorporate dynamical (geometric) corrections to the brane geometry.

\sp
\noindent
Under $r\leftrightarrow -r$, the emergent geometries correspond to an effective anti $D_3$-brane. In a global scenario, i.e. both brane and its anti-brane together, two conserved charges associated with the emergent metric components $G_{tr}$ and $G_{r\phi}$ in eq.(\ref{RN-2}) may seen to disappear. For functions $(g_1,g_2,g_3)$, the invariant curvatures in the classical geometry (\ref{RN-2}) are computed in Einstein's gravity. They may be given by

\vspace{-.15in}
$${}\;\;\;\;\ R={{(Mp)^2}\over{r^4\sin^4 \t }}\ g_1(r,\theta,M,p)\ ,\qquad\qquad {}$$
\vspace{-.2in}
$$R_{\mu\nu}R^{\mu\nu}= {{M^2}\over{32r^8\sin^8\t}}\ g_2(r,\theta,M,p)$$
\vspace{-.2in}
$${\rm and}\quad\;\ R_{\mu\nu\lambda\rho}R^{\mu\nu\lambda\rho}= {{M^2}\over{r^8}}\ g_3(r,\theta,M,p)\ .\qquad\qquad {}$$

\noindent
In addition to a curvature singularity at $r\rightarrow 0$, a numerical analysis re-assures that the curvature blows up at the horizon $r\rightarrow M$ for a large $M$ and an arbitrary $p$. However, a consistent dynamical description in a global scenario enforces a constraint between the parameters $(p,Q_e,Q_m)$ in the emergent geometries (\ref{RN-2}). For $p$$>$$0$ and $(Q_e^2\pm Q_m^2)=p^2 \sin^2\theta$, the line-elements (\ref{RN-2}) for a $(D{\bar D})_3$-pair is given by 
\bea
ds^2&=&-\left (1-{{M^2}\over{r^2}}\right ) dt^2 + \left ( 1 -{{M^2}\over{r^2}}\right )^{-1} dr^2 + r^2 d\Omega^2\nonumber\\
&&\;\ -\ {{2M^4\sin^2\theta}\over{r^2}} \Omega_{\phi}^{\pm}\ dt d\phi \ \pm\ {{Q_e^2}\over{r^4}} ds^2(g^{\pm}_{\mu\nu})\ .\label{RN-3}
\eea
Interestingly, the line-element in the quantum correction describes a Minkowski space-time and is given by
$$ds^2(g^{\pm}_{\mu\nu})=\left ( -dt^2 +\ dr^2 \pm\ r^2 d\Omega^2\right )\ .\qquad\qquad\qquad\qquad {}$$
It is interesting to note that the modified Schwarzschild geometries undergo rotations. Their angular velocities are computed at the horizon $r_s=M$ for $M>|Q_e|$ to yield
\vspace{-.1in}
\noindent
$$\Omega_{\phi}^{\pm}= {{\sqrt{Q_e^2\pm Q_m^2}}\over{M^3 \sin\theta}}\ .$$ 
The dependence of $\Omega_{\phi}^{\pm}$ on the non-linear $U(1)$ charges is remarkable. It re-confirms the quantum corrections, leading to a new characteristic feature of rotation, into a typical Schwarzschild black hole on an emergent brane. The presence of angular momentum is intrinsic to the frame-work underlying a dynamical torsion.

\sp
\noindent
On the other hand, the near horizon geometry of the Schwarzschild black hole (\ref{RN-3}), with ($\Omega_{\phi}^-$ and $-Q_e^2$) choice, may be expressed for a sensible quantum correction under a flip of light cone at the horizon. In the case, the quantum geometry undergoes a geometric tunneling to lead to an asymptotic AdS with a correct quantum correction. It is given by 
\bea
ds^2&=&-{{r^2}\over{b^2}} dt^2+ {{b^2}\over{r^2}} dr^2 + r^2 d\Omega^2\\ \nonumber && \qquad- {{2b^4\sin^2\theta}\over{r^2}} \Omega_{\phi}^-\ dt d\phi\ 
+\ {{Q_e^2}\over{r^4}} ds^2(g^+_{\mu\nu})\ .\label{RN-4}
\eea
The analysis re-assures the fact that an electric charge incorporates a quantum correction to deform a Schwarzschild geometry in the frame-work. Interestingly the quantum corrections to Schwarzschild black hole with $\Omega_{\phi}^+$ and $+Q_e^2$ in (\ref{RN-3}) and an asymptotic AdS in (\ref{RN-4}) are equal and are associated with a flat metric. The quantum corrections dominate for $|Q_e|$$>$$r$. However for $Q_e$$<$$r$, the global brane geometries shall be seen to describe semi-classical black holes established in Einstein's gravity. In absence of a magnetic monopole $\Omega_{\phi}^+\rightarrow\Omega_{\phi}^-$. A priori, two inequivalent classical vacua receive precisely the same quantum correction in presence of an electric point charge in the frame-work. Interestingly for $p=0$, both the Schwarzschild black holes (\ref{RN-2}) tunnel to an asymptotic AdS geometry with $(-)$ve and $(+)$ve quantum corrections:
\be
ds^2_q=\mp {1\over{r^4}}\Big ( Q_e^2 [-dt^2+dr^2]\ +\ Q_m^2\ r^2 d\Omega^2 \Big ) \ .\label{RN-41}
\ee

\section{Emergent Reissner-Nordstrom and Schwarzschild black holes}
In this section, we obtain a Reissner-Nordstrom and a Schwarzschild emergent black holes on an effective $D_3$-brane underlying a $U(1)$ gauge dynamics of two form and an one form. We compute the black hole seiberg-witten entropy at the horizon of both the brane geometries. Interestingly, our analysis for emergent geometries re-confirm $S_{\rm Sch}>S_{\rm RN}$, which is in agreement with the Einstein's gravity.

\sp
\noindent
In the limit $r$$>$$>$$b$$>$$(2\pi\alpha')^{1/2}|Q_e|>1$, with a fixed $\alpha'$, the semi-classical geometries (\ref{RN-1}) on an effective $D_3$-brane is worked out. Within the regime, the emergent quantum geometries become 
\bea
&&ds^2=-\left (1-{{M^2}\over{r^2}} \pm {{Q_e^2}\over{r^4}}\right ) dt^2 + {{2M^2}\over{r^2}}\ dt dr {}\nonumber\\
&&+\left ( 1 -{{M^2}\over{r^2}} \mp {{Q_e^2}\over{r^4}}\right )^{-1} dr^2 -\ {{2Mp \sin^2 \theta}\over{r^2}}\left (dt + dr\right )d\phi\nonumber\\
&&+\left ( 1 + {{{p^2 \sin^2\theta}\mp Q_m^2}\over{r^4}}\right )\ r^2 d\Omega^2\ .\label{RN-5}
\eea
A priori, the emergent geometries may be viewed as an orthogonal combination of a Reissner-Nordstrom and a Schwarzschild black holes.
The semi-classical Reissner-Nordstrom geometry may be projected out using a generalized ($2\times2$) matrix $N$ prescription \cite{D4-torsion} containing the longitudinal components of metric in lorentzian signature. The matrix is given by
\bea
&&2N={\tilde N}=\left(\begin{array}{ccc}
-{{G}_{tt}^+}&&{{G}_{rr}^+}\\
{{G}_{rr}^-}&&{-{G}_{tt}^-}
\end{array}\right )\qquad {\rm and} \qquad {}\nonumber\\
{\rm where}&& G_{tt}^{\pm}= \left ( 1 -{{M^2}\over{r^2}}\pm {{Q_e^2}\over{r^4}}\right )=\left ( G_{rr}^{\pm}\right )^{-1}\ .\label{RN-6}
\eea
\bea
{\rm Then}\ {\tilde N}\left ( \begin{array}{c}
1\\
0
\end{array}\right)=\left( \begin{array}{c}
{-{G}_{tt}^+}\\
{{G}_{rr}^-}
\end{array}\right),\ {\tilde N}\left( \begin{array}{c}
0\\
1
\end{array}\right)=\left( \begin{array}{c}
{{G}_{rr}^+}\\
{-{G}_{tt}^-}
\end{array}\right),&&\nonumber\\
{\tilde N}^{-1}\left( \begin{array}{c} 1\\ 0
\end{array}\right)=\left( \begin{array}{c}
{G_{tt}^-}\\ {G_{rr}^-}
\end{array}\right),\ {\tilde N}^{-1}\left( \begin{array}{c}
0\\
1
\end{array}\right)=\left( \begin{array}{c}
{G_{rr}^+}\\
{G_{tt}^+}
\end{array}\right).&&\nonumber
\eea 
The determinant of the generalized matrix, in absence of electric charges, at the Schwarzschild horizon ensures ($\det N=-1$) a discrete transformation between the two sets of emergent geometries. The projections, of the matrix $N$, ensure the mixed emergent brane geometries (\ref{RN-5}), while the projections of the inverse matrix $N^{-1}$ on the same yield new interesting euclidean geometries on an effective $D_3$-brane. They are
\bea
ds^2&=&\left (1-{{M^2}\over{r^2}} \pm {{Q_e^2}\over{r^4}} \right ) dt^2+ 
\left ( 1 -{{M^2}\over{r^2}} \pm {{Q_e^2}\over{r^4}}\right )^{-1} dr^2\nonumber\\
&+&\left ( 1 + {{{p^2 \sin^2\theta}\mp Q_m^2}\over{r^4}}\right ) r^2 d\Omega^2
+\ {{2M^2}\over{r^2}}\ dt dr\nonumber\\ 
&&\qquad\qquad\qquad -\ {{2Mp \sin^2 \theta}\over{r^2}}\left (dt + dr\right )d\phi\ .\label{RN-10}
\eea
The global brane geometries,
for $Q_m=p\sin \theta$, become
\bea
ds^2&=&\left (1-{{M^2}\over{r^2}} \pm {{Q_e^2}\over{r^4}} \right ) dt^2+ 
\left ( 1 -{{M^2}\over{r^2}} \pm {{Q_e^2}\over{r^4}}\right )^{-1} dr^2\nonumber\\
&&\qquad\quad +\ f^{\mp}\ r^2 d\Omega^2\ -\ {{2{\cal Q}_m}\over{r^2}}\ dt (\sin \theta d\phi)\ , \label{RN-11}
\eea
\vspace{-.2in}
$${\rm where}\quad f^-=1,\; f^+= \left ( 1+{{2Q_m^2}\over{r^4}}\right )\;\ {\rm and}\; {\cal Q}_m=MQ_m\ .$$ 
The geometry, with ($+Q_e^2$, $f^-$), in (\ref{RN-11})
describes an emergent Reissner-Nordstrom black hole in presence of a non-linear magnetic charge ${\cal Q}_m$. Using the area law, the black hole entropy 
is computed at the event horizon $r_+$ to yield
$$S_{\rm RN}\approx \pi M^2\left ( 1 -{{Q_e^2}\over{M^4}}\right )\ .$$ 
The other geometry, with ($-Q_e^2$, $f^+$), in (\ref{RN-11}) a priori describes an emergent Schwarzschild geometry with a re-defined black hole mass ${\tilde M}$ and a charge ${\cal Q}_m$. It is given by
\bea
ds^2&=&\left (1-{{{\tilde M}^2}\over{r^2}} \right ) dt^2+ 
\left ( 1 -{{{\tilde M}^2}\over{r^2}}\right )^{-1} dr^2\nonumber\\ 
&+&\left ( 1 + {{2Q_m^2}\over{r^4}}\right ) r^2 d\Omega^2 - {{2{\cal Q}_m}\over{r^2}} dt (\sin \theta d\phi)\ . \label{SCZ-1}
\eea
Explicitly the mass, defined at the horizon, is given by $${\tilde M}= M\Big (1 + {{Q_e^2}\over{M^4}}\Big )^{1/2}\ .$$
The modified mass turns out to be a conserved quantity due to its association with an electric charge $Q_e$ in the frame-work. The correction in ${\tilde M}$ is intrinsic to the emergent geometry on an effective $D_3$. The charge $Q_e$ incorporates dynamics into the emerging gravity on a brane. Importantly,
 the emerging black hole entropy at its horizon, $r=r_s$, is computed to yield
$$S_{\rm Sch}\approx \pi M^2\left (1 + {{Q_e^2}\over{M^4}}\right ) + {{2\pi Q_m^2}\over{M^2}}\left (1 - {{Q_e^2}\over{M^4}}\right )\ .$$ 
It re-assures that $S_{\rm Sch}>S_{\rm RN}$ for $Q_m= p \sin\theta$. In absence of a magnetic monopole, the emergent geometry on a brane precisely describes a typical Schwarzschild black hole. In the case the emergent black hole entropy, at its horizon, becomes $S_{\rm Sch}=\pi{\tilde M}^2$ and may be identified with the entropy of a Schwarzschild black hole in Einstein's gravity. 

\sp
\noindent
Interestingly, the invariant scalar curvatures in Einstein's gravity computed for the complete emergent Schwazschild geometry (\ref{SCZ-1}) on a brane shows that the Ricci scalar $R=0$ in absence of a magnetic monopole. In the case, the space-time curvature blows up in the limit $r\rightarrow 0$. 
For a generic black hole on a brane, $i.e.\ Q_m\neq 0$, the curvature singularities are a priori seen to be at $r\rightarrow 0$ and at the horizon $r_h=M$. However for $|Q_m|$$>$$>$$|M|$, both the singularities in all the three curvature scalars ($R$, $R_{\mu\nu}R^{\mu\nu}$ and $R_{\mu\nu\lambda\rho}R^{\mu\nu\lambda\rho}$) may seen to be covered by an event horizon $r_S=M$.

\sp
\noindent
On the other hand, for $p=0$, the emergent global brane geometry with $+Q_e^2$ and $-Q_m^2$ in (\ref{RN-10}) may be expressed with a real time. 
For higher dimensional Reissner-Nordstrom, see ref.\cite{gao-zhang}. In the case, the Reissner-Nordstrom black hole is given by
\bea
&&ds^2=-\left (1-{{M^2}\over{r^2}} + {{Q_e^2}\over{r^4}}\right ) dt^2\qquad\qquad\qquad\qquad\qquad {}\nonumber\\
&&\ + \left ( 1 -{{M^2}\over{r^2}} + {{Q_e^2}\over{r^4}}\right )^{-1} dr^2 + \left ( 1 - {{Q_m^2}\over{r^4}}\right ) r^2 d\Omega^2\ .\label{RN}
\eea 
The invariant curvature scalars in Einstein's gravity, for an emergent black hole on a brane, are computed. The space-time curvature blows up independently in both the limits $r\rightarrow 0$ and $r\rightarrow\sqrt{Q_m}$. For a large $M$, $i.e.\ M^2$$>$$>$$|Q_m|$ and $M^2$$>$$>$$|Q_e|$, both the curvature singularities may seen to be covered by an inner horizon $r_-$ in eq.(\ref{RN}). 
The precise geometric correspondence between an emergent black hole, underlying a two form $U(1)$ gauge theory on a $D_3$-brane, and the Reissner-Nordstrom black hole in Einstein's gravity is remarkable. It further re-assures the fact of vanishing torsion in the gauge choice for a two form in the frame-work. Contrary to Einstein's gravity, the black hole mass $M^2$ in the emergent Reissner-Nordstrom geometry is primarily sourced by a $B_{\mu\nu}$ potential on a $D_3$-brane. It may provoke thought to explore a $B_{\mu\nu}$ potential, at the expense of a metric potential, in an effective theory of gravity.

\sp
\noindent
The black hole entropy in the case becomes 
$$S_{\rm RN}\approx \pi M^2 - {{\pi Q_e^2}\over{M^2}}\left ( 1 + {{4Q_m^2}\over{M^4}}\right )\ .$$ 

\noindent
The remaining emergent geometry in (\ref{RN-10}) for ($-Q_e^2$, $f^+$) describes a Schwarzschild black hole
\bea
&&ds^2=-\left (1-{{{\tilde M}^2}\over{r^2}} \right ) dt^2+ 
\left ( 1 -{{{\tilde M}^2}\over{r^2}}\right )^{-1} dr^2\qquad\qquad {}\nonumber\\
&&\qquad\qquad\qquad\qquad\quad\quad\ \ + \left ( 1 + {{Q_m^2}\over{r^4}}\right ) r^2 d\Omega^2\ .\label{SCZ-2}
\eea

\sp
\noindent
On the other hand, a magnetic monopole in an emergent Reissner-Nordstrom reduces its $S^2$-radius and hence shrinks its geometry, while it enhances the $S^2$-radius in a Schwarzschild geometry on an effective $D_4$-brane on $S^1$. A priori, a magnetic monopole does not significantly modify the black hole except for a conformal factor in its spherical geometry. However, analysis reveals that the dynamical aspects of Einstein's gravity may be viewed via a magnetic monopole in the emergent Schwazschild black hole (\ref{SCZ-2}) on a brane.
The space-time curvatures are computed for an emergent Schwarzschild geometry (\ref{SCZ-2}) and they are:
$$R=0\; ,\quad R_{\mu\nu}R^{\mu\nu}={1\over{r^8\left (Q_m^2+r^4\right )^4}}=R_{\mu\nu\lambda\rho}R^{\mu\nu\lambda\rho}\ .$$
Hence, a brane-world Schwarzschild black hole possesses a curvature singularity at $r\rightarrow 0$ and a coordinate singularity at $r_S={\tilde M}$.

\sp
\noindent
The entropy for an emergent Schwarzschild black hole is computed to yield:
$$S_{\rm Sch}\approx \pi M^2 \left ( 1 + {{Q_e^2}\over{M^4}}\right )+ {{\pi Q_m^2}\over{M^2}}\left ( 1 - {{Q_e^2}\over{M^4}}\right )\ .$$ 
The computed entropy further confirms $S_{\rm Sch}>S_{\rm RN}$ for $p=0$. The result is in agreement with the black holes in Einstein's gravity.

\sp
\noindent
In other words, the electro-magnetic charges incorporate local degrees into the emergent global brane geometries leading to a typical Reissner-Nordstrom black hole (\ref{RN}) and a typical Schwarzschild black hole (\ref{SCZ-2}) in the frame-work. The local degrees of the geometric ${\cal A}_{\mu}$ field in the $U(1)$ gauge theory on a $D_3$-brane may precisely be identified with the dynamical metric tensor field in Einstein's gravity. Importantly, the emergent metric potential in Reissner-Nordstrom black hole and Schwarzschild black hole further re-assures the presence of an extra fifth compact dimension in the frame-work.

\sp
\noindent
The emergent Schwarzschild and Reissner-Nordstrom geometries may seen to be influenced by a generic torsion underlying a $D_4$-brane \cite{D4-torsion}. The two form ansatz leading to a generic torsion in five dimensions may be reviewed to enhance our understanding on the four dimensional vacua in Einstein's gravity. The gauge field ansatz on a $D_4$-brane were
\bea
&&{\tilde B}_{t\theta}={\tilde B}_{r\theta}={b\over{\sqrt{2\pi\alpha'}}}\ ,\qquad\qquad\qquad {}\nonumber\\
&&{\tilde B}_{\theta\phi}={{p}\over{\sqrt{2\pi\alpha'}}} \sin^2\theta \cos \psi \nonumber\\
{\rm and}&&{\tilde B}_{\psi\theta} ={{\tilde p}\over{\sqrt{2\pi\alpha'}}}\sin^2\theta\cot\psi \ .\label{RN-14}
\eea
The nontrivial geometric torsion, sourced by the two form, was worked out to yield
\bea
{{\cal H}_{\psi\phi}}^{\theta}&=&\ {{p}\over{r^2}}\sin^2\theta \sin \psi\nonumber\\
{\rm and}\;\; {{\cal H}_{\psi\phi}}^t&=&-{{\cal H}_{\psi\phi}}^r={1\over{\sqrt{2\pi\alpha'}}} {{bp}\over{r^2}} \sin^2\theta \sin \psi\ .\label{RN-15}
\eea
For $\psi=n\pi$ (integer $n$), the ansatz for a two form on a $D_4$-brane precisely reduces to that on a $D_3$-brane (\ref{gauge-8}). Though the local degree of two form freezes on a $D_3$-brane, the emerging gauge field retains the dynamical aspects of a $D_3$-brane.
\section{Concluding remarks}
In the article, we have revisited a generalized curvature formulation underlying a torsion dynamics in an effective $D_4$-brane on $S^1$.
Interestingly, the gauge choice for a local torsion on a $D_4$-brane is shown to describe torison free geometry when considered on $S^1$. As a result, the two form behaves like a background potential and has been argued to source the mass of a Schwarzschild black hole in a limit. We have obtained, a Reissner-Nordstrom and a Schwarzschild, emergent black holes in a gauge choice for a generalized five dimensional curvature on $S^1$.  The electric charge has been shown to incorporate quantum geometric corrections to the Schwarzschild and asymptotic AdS emerging geometries. The magnetic charge has been argued to shrink the radius of $S^2$ for an emergent Reissner-Nordstrom black hole. On the other hand, $S^2$-radius has been shown to grow in an emergent Schwarzschild black hole in presence of a magnetic monopole. 

\sp
\noindent
The fact that both the emergent black holes (\ref{RN-11}) in the frame-work may precisely be identified with the Schwarzschild vacua of different masses ${\tilde M}\rightarrow M\sqrt{1 \pm {{Q_e^2}/{M^4}}}$ in absence of a magnetic monopole is remarkable. Firstly, it ensures Einstein's gravity in $4D$ from a generalized curvature underlying a geometric torsion dynamics on a $D_4$-brane. Secondly our analysis based on a generalized curvature connects two different vacuum solutions, $i.e.$  Schwarzschild and Reissner-Nordstrom black holes, in Einstein's gravity. The geometric transition or tunneling between the two vacua is enforced by the quantum fluctuations in a two form $U(1)$ gauge theory on a $D_4$-brane and needs further attention in recent research.

\section*{Acknowledgments}
A.K.S. and S.S. respectively acknowledge the Council of Scientific and Industrial Research (CSIR) and University Grants Commission (UGC) for their fellowship. The work of S.K. is partly supported by a research grant-in-aid under the Department of Science and Technology (DST), Govt.of India.

\def\anp{Ann. of Phys.}
\def\prl{Phys.Rev.Lett.}
\def\prd#1{{Phys.Rev.}{\bf D#1}}
\def\jhep{JHEP\ {}}{}
\def\cqg#1{{Class. \& Quant.Grav.}}{}
\def\plb#1{{Phys. Lett.} {\bf B#1}}
\def\npb#1{{Nucl. Phys.} {\bf B#1}}
\def\mpl#1{{Mod. Phys. Lett} {\bf A#1}}
\def\ijmpa#1{{Int.J.Mod.Phys.}{\bf A#1}}
\def\mpla#1{{Mod.Phys.Lett.}{\bf A#1}}
\def\rmp#1{{Rev. Mod. Phys.} {\bf 68#1}}
\def\grg{Gen.Relativ.Gravit.}{}

\end{document}